\documentclass{PoS}

\title{Inclusive Jet Measurements in Longitudinally Polarized proton-proton Collisions at STAR}

\ShortTitle{STAR Inclusive Jet Measurements}

\author{\speaker{Zilong Chang for the STAR collaboration}
	\\
        Brookhaven National Laboratory\\
        E-mail: \email{zchang@bnl.gov}}

\abstract{Jet production in high energy proton-proton ($p+p$) collisions is dominated by hard QCD subprocesses such as gluon-gluon and quark-gluon scatterings. Therefore $p+p$ is an effective tool to probe the internal distribution of gluons in the proton. The STAR Collaboration at the Relativistic Heavy Ion Collider (RHIC) is using longitudinally polarized $p+p$ collisions at the center of mass energies, $\sqrt{s} = $ 200 and 510 GeV, to study the cross-section and double spin asymmetries, $A_{LL}$, of inclusive jet and di-jet production. The measurements at $\sqrt{s} = $ 200 GeV showed that the jet cross-sections are well described by the next-to-leading-order perturbative QCD calculations after underlying event and hadronization corrections. The previous 2009 $\sqrt{s} = $ 200 GeV inclusive jet asymmetry $A_{LL}$ measurement at pseudo-rapidity $|\eta| < $ 1.0 showed the first experimental evidence of a non-zero gluon polarization for partonic momentum fraction, $x > 0.05$. The inclusive jet $A_{LL}$ measurement at $\sqrt{s} = $ 510 GeV allows to explore the gluon polarization to smaller $x \sim 0.02$. The 2012 inclusive jet $A_{LL}$ and cross-section measurements at $\sqrt{s} =$ 510 GeV, the techniques used in the jet analysis including underlying event correction, and future perspectives related to jet measurements at STAR will be presented.
}

\FullConference{XXVI International Workshop on Deep-Inelastic Scattering and Related Subjects (DIS2018)\\
		16-20 April 2018\\
		Kobe, Japan}

\begin{document}


\section{Introduction}
\begin{figure}[h]
\centering
\includegraphics[scale=0.6]{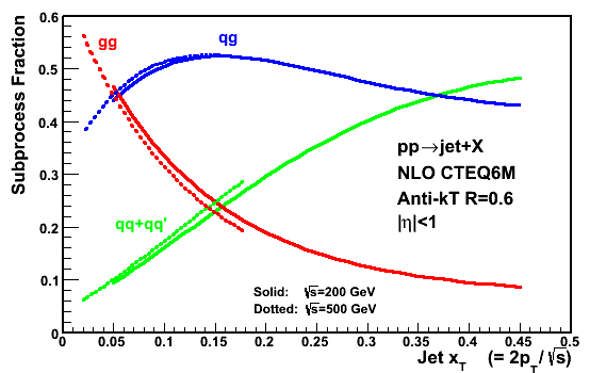}
\caption{Fraction of jet production due to three sub-processes, $gg$, $qg$ and $qq$ in $p+p$ collisons at $\sqrt{s}
= $ 200 and 500 GeV \cite{nlosubproc}.}
\label{fig:procratios}
\end{figure}

In high energy $p+p$ collisions, parton scatterings through hard QCD processes produce a group of collimated energetic particles, so called jets. At the RHIC kinematics, the gluon-gluon ($gg$)
and quark-gluon ($qg$) processes dominate the jet production over the quark-quark ($qq$) process, see Figure \ref{fig:procratios} \cite{nlosubproc}. Therefore the STAR experiment measures jets to gain more detailed insights into the gluon distribution inside the proton. 

RHIC is a 2.4 mile circular collider capable of colliding both transversely and longitudinally polarized proton beams up to the center of mass energy of $\sqrt{s} = $ 510 GeV. The featured detectors at STAR include the Time Projection Chamber (TPC), the Barrel Electromagnetic Calorimeter (BEMC), and the Endcap Electromagnetic Calorimeter (EEMC). The TPC is the major tracking system to measure charged particles in the acceptance $-1.3< \eta < 1.3$ and $0 < \phi < 2\pi$. Both BEMC and EEMC are made of lead-scintillator stacks with full azimuthal coverage and psuedo-rapdity coverage, $-1.0< \eta <1.0$ and $1.1< \eta <2.0$, respectively. They are also divided into multiple $1 \times 1$ jet patches in $\Delta \eta$ and $\Delta \phi$ space for triggering on high energy jet events. The Vertex Position Detector (VPD), Zero Degree Calorimeter (ZDC) and Beam Beam Counter (BBC) are used for relative luminosity measurements \cite{rhic}.

The double-spin asymmetry, $A_{LL}$, for inclusive jets, is sensitive to the gluon polarization in the proton, which is poorly constrained by Deep Inelastic Scattering (DIS) experiments. The STAR 2009 inclusive jet $A_{LL}$ at $\sqrt{s} = $ 200 GeV provided the first evidence of positive gluon polarization at $x > 0.05$ \cite{run9jet,dssv,nnpdf}. At higher $\sqrt{s} = $ 510 GeV, the inclusive jet $A_{LL}$ will constrain the gluon polarization at lower $x \sim$ 0.02. Di-jet measurements are complementary as they unfold the kinematics of both scattering partons and therefore provide more details of the shape of the polarized gluon distribution function.

\section{Jet Reconstruction}
At STAR, jets are reconstructed from the charged tracks measured by the TPC and the neutral-particle energy deposited in the BEMC and EEMC. A midpoint cone algorithm \cite{midpnt} with a cone radius of $0.7$ was used in the early 2006 $\sqrt{s} = 200$ GeV data analysis. In the later analysis of the 2009 $\sqrt{s} = 200$ GeV data, the anti-$k_T$ algorithm \cite{antikt} with the jet parameter $R = 0.6$ was used. For the 510 GeV data, a smaller jet parameter value of $R = 0.5$ was chosen to account for the increased soft diffuse background and more pile-up events in the higher center-of-mass energy.

\section{Underlying Event Correction}
\begin{figure}[h]
\centering
\includegraphics[scale=0.6]{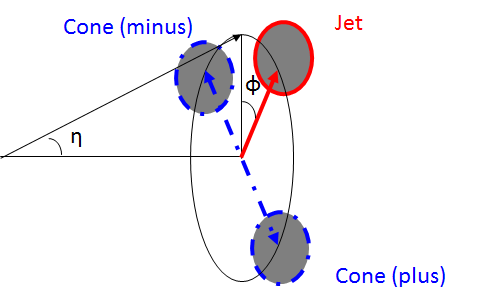}
\caption{The two off-axis cones centered at $\pm \frac{\pi}{2}$ away in $\phi$ and at the same $\eta$ relative to a given
jet.}
\label{cones}
\end{figure}

Underlying events, also known as soft and random multiple parton interactions, present an additional challenge for the jet analysis. A technique of applying underlying event corrections to the jet transverse momentum, $p_T$, adapted from the ALICE experiment \cite{cones}, is used in the recent STAR jet analyses. The underlying event energy density is obtained from the average energy density,
$\hat{\rho}$, deposited in the two off-axis cones centered at the same $\eta$ of the jet, but $\pm \frac{\pi}{2}$ away from the jet in the azimuthal direction, thus named off-axis method, see Figure \ref{cones}. The jet transverse momentum correction, $dp_T$, is taken as the average cone density, $\hat{\rho}$, times the jet area, $A$. This procedure allows to remove underlying event contribution on a jet-by-jet basis and to sample the $\eta$ dependence of the underlying event contribution.

\begin{figure}[h]
\centering
\includegraphics[scale=0.6]{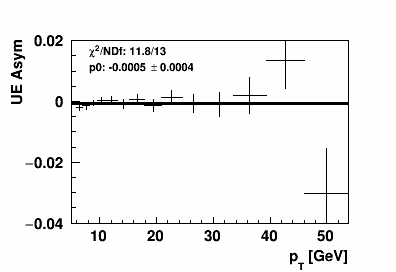}
\caption{Underlying event correction $dp_T$ asymmetries, $A_{LL}^{dp_T}$ vs. underlying event corrected jet $p_T$ with a constant fit, in STAR 2012 510 GeV
data.}
\label{ueasym}
\end{figure}

Furthermore it permits to study the spin-dependence of the underlying event contribution to the jet in longitudinally polarized $p+p$ collisions. An underlying event correction $dp_T$ asymmetry is defined as,
\begin{equation}
A_{LL}^{dp_T} = \frac{1}{P_AP_B} \frac{(<dp_T>^{++} +<dp_T>^{--}) - (<dp_T>^{+-} +<dp_T>^{-+})}{(<dp_T>^{++}
+<dp_T>^{--}) + (<dp_T>^{+-} +<dp_T>^{-+})},
\end{equation}
where $P_{A,B}$ is the polarization of beam A, B, $<dp_T>^{++,+-,-+,--}$ is the average underlying event correction for a given spin state, $++$, $+-$, $-+$, or $--$. The $dp_T$ asymmetry is investigated in each of the jet $p_T$ bins. In the STAR 2012 510 GeV longitudinally polarized $p+p$ data, it is consistent with zero, see Figure \ref{ueasym}, which leads to a small underlying event contribution to the jet $A_{LL}$.

\section{Data and Embedding Comparison}
For data collected in each year, a set of embedded events is generated with PYTHIA \cite{pythia} as the event generator. The detector responses are simulated by GEANT \cite{geant}. At the last step, the simulated responses are embedded into zero-bias events from data. The intermediate parton and particle records are saved from PYTHIA, which allows to reconstruct parton and particle jets and to compare them with detector jets reconstructed from the embedded detector responses. 

\begin{figure}[h]
\centering
\includegraphics[scale=0.4]{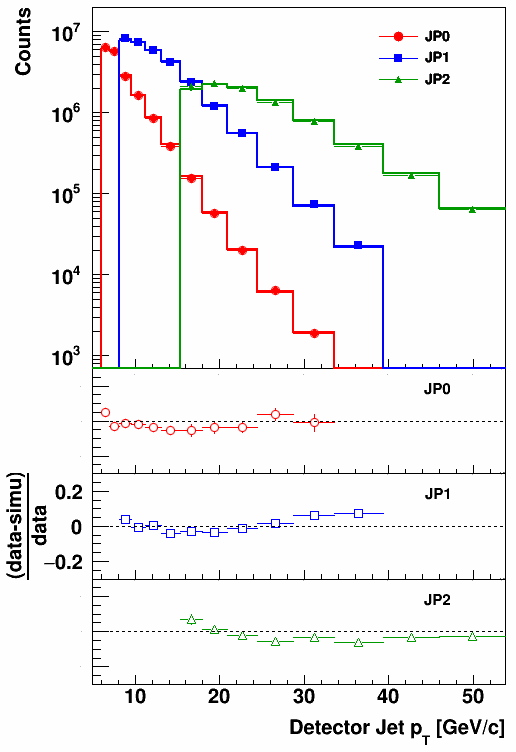}
\caption{The Jet spectra comparison for three jet patch triggered events in STAR 2012 510 GeV data.}
\label{jetpt}
\end{figure}

The PYTHIA simulation based on the Perugia 2012 tune \cite{perugia} shows increased $\pi^{\pm}$ yields at low $p_T$ compared to the previously published STAR results \cite{starpi1,starpi2}. This can be explained by excess of contributions from soft scattering processes, such as multiple parton interactions. To reduce this contribution, the exponent factor, $P_{90}$, controlling the center-of-mass energy dependent cut-off in transverse momentum, $p_{T,0}$, is decreased from 0.24 to 0.213. This leads to 7\% increase in the $p_{T,0}$ for $\sqrt{s} = 510$ GeV, see Equations \ref{eq:sigma} and \ref{eq:pt0}. With this modification, the jet spectra show good agreement between data and embedding for the 2012 510 GeV data, see Figure \ref{jetpt}, and the embedding sample reproduces our data well.

\begin{equation}
\sigma \sim \frac{1}{(p_T^2+p_{T,0}^2)^2}
\label{eq:sigma}
\end{equation}

\begin{equation}
p_{T,0} = p_{T,ref} \times (\frac{\sqrt{s}}{\sqrt{s_{ref}}})^{P_{90}}
\label{eq:pt0}
\end{equation}

\section{Future and Conclusion}
STAR is proposing to install a Forward Calorimeter System (FCS), which consists of an electromagnetic calorimeter and a hadron calorimeter, and a Forward Tracking System (FTS) in the 2020s. This will enable STAR to measure di-jets with one or both jets at rapidities of $2.8 < \eta < 3.7$. The measured di-jet $A_{LL}$ in this region will access the gluon polarization down to $x \sim 10^{-3}$ \cite{upgrade}.

In summary, the inclusive jet and di-jet measurements are an integral part of the STAR physics program to study the internal structure of the proton and the famous proton spin puzzle. Jet production has been well understood over the years. With the new $\sqrt{s} = 200 $ GeV data from 2015, the previous statistical uncertainties will be reduced by a factor of two. The combined inclusive jet $A_{LL}$ will further constrain the gluon polarization in the proton. A lot of more exciting jet results will follow in the near future.

\end{document}